\documentclass[aps,prl,twocolumn,preprintnumbers,amsmath,amssymb]{revtex4}
\usepackage{graphicx}
\usepackage{amsmath}
\usepackage{amsfonts}
\usepackage{epsfig}
\usepackage{bm}
\usepackage{color}

\begin{document}

\title{Superconductivity in the doped quantum spin liquid on the triangular lattice}
\author{Hong-Chen Jiang}
\email{hcjiang@stanford.edu}
\affiliation{Stanford Institute for Materials and Energy Sciences, SLAC and Stanford University, Menlo Park, California 94025, USA}
\date{\today}

\begin{abstract}
Broad interest in quantum spin liquid (QSL) phases was triggered by the notion that they can be viewed as insulating phases with preexisting electron-pairs, such that upon light doping they might automatically yield superconductivity. Yet despite intense efforts, definitive evidence is lacking. We address the problem of a lightly doped QSL through a large-scale density-matrix renormalization group study of the $t$-$J$ model on the triangular lattice with a small but non-zero concentration of doped holes. The ground state is consistent with a Luther-Emery liquid with power-law superconducting and charge-density-wave correlations associated with partially-filled charge stripes. In particular, the superconducting correlations are dominant on both four-leg and six-leg cylinders at all hole doping concentrations. Our results provide direct evidences that doping a QSL can naturally lead to robust superconductivity.
\end{abstract}
\maketitle

Quantum spin liquids (QSLs) are exotic phases of matter that exhibit a variety of novel features associated with their topological character.\cite{Anderson1973,Balents2010,Savary2016,Zhou2017,Broholm2019} 
The simplest QSLs known theoretically are characterized by topological order and support fractional excitations such as spinons, which carry the spin-1/2 but not the charge of the electron. The close relationship between a QSL and a superconductor suggests that doping it might naturally lead to superconductivity.\cite{Broholm2019,Anderson1987,Kivelson1987,Rokhsar1988,Laughlin1988,Laughlin1988S,Wen1989,Emery1997,Senthil2005,Lee2007,Lee2006,Fradkin2015}  Indeed, this is the original dream of the resonating valence bond (RVB) theory as a mechanism for high temperature superconductivity.\cite{Anderson1987} However, despite intense efforts during past three decades, definitive evidence showing that doping QSLs gives rise to superconductivity is lacking.\cite{Broholm2019,Lee2006,Fradkin2015,Kelly2016,Jiang2017} This is partially due to the fact that the realization of QSLs is a great challenge to physicists, where candidate materials and systems are rare and somewhat controversial.\cite{Balents2010,Savary2016,Zhou2017,Broholm2019} 

\begin{figure}
  \includegraphics[width=\linewidth]{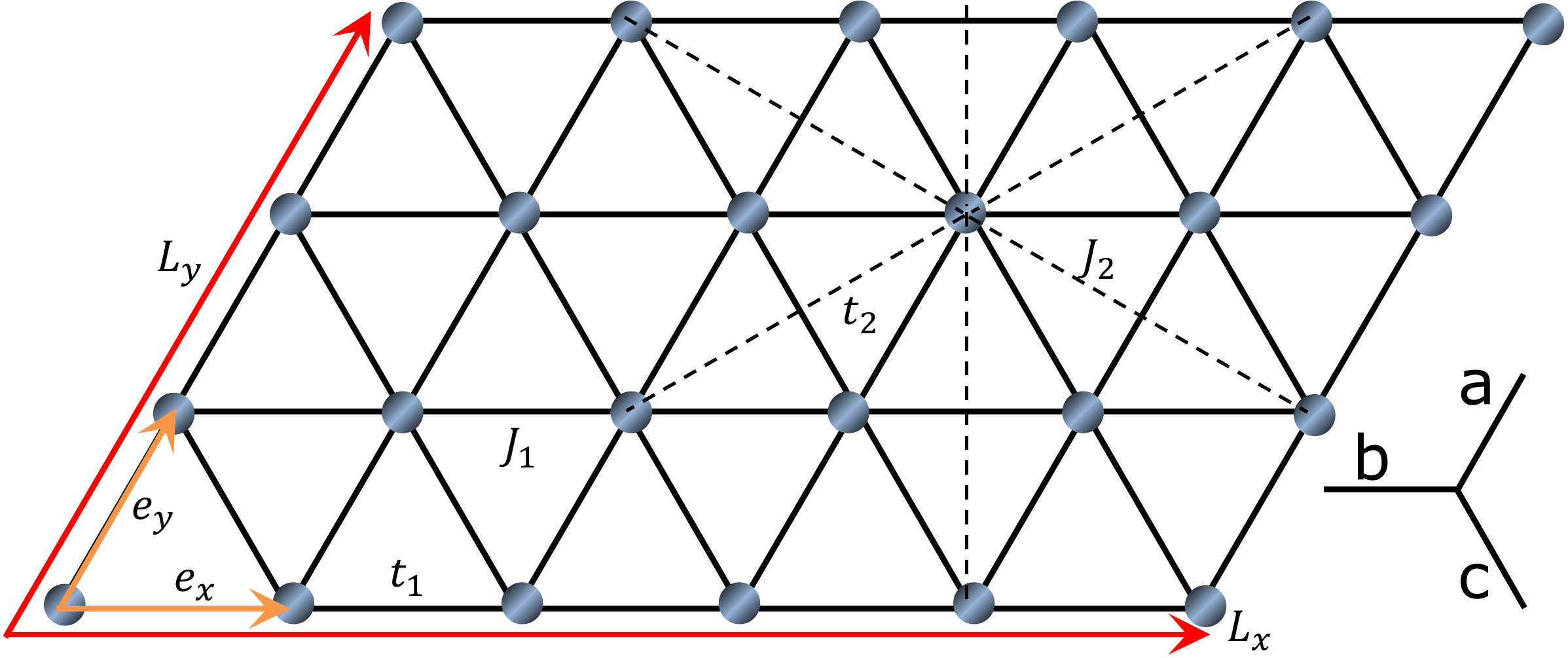}
  \caption{The $t$-$J$ model on the triangular lattice. Periodic and open boundary conditions are imposed respectively along the directions specified by the lattice basis vectors $\mathbf{e}_y$ and $\mathbf{e}_x$. $L_x$ and $L_y$ are the number of sites in the $\mathbf{e}_x$ and $\mathbf{e}_y$ directions. $t_1$ ($t_2$) and $J_1$ ($J_2$) are hopping integral and spin exchange interactions between NN (NNN) sites. The electrons live at the vertices (filled circles), and a, b and c denote the three different bonds.}\label{Fig:Lattice}
\end{figure}

Faced with these challenges, quasi-one-dimensional systems such as cylinders (depicted in Fig.\ref{Fig:Lattice}) have become an important starting point to resolve this problem. The cylinder can be viewed as one-dimensional (1D) but has essential degrees of freedom that allow for two-dimensional (2D) characteristics to emerge. Moreover, it can be accurately studied using the density-matrix renormalization group (DMRG) numerical procedure\cite{White1992,Schollwock2011,Stoudenmire2012}. The Luther-Emery (LE) liquid\cite{Luther1974} is a 1D analog of a superconductor which can be realized in quasi-1D systems such as cylinders.\cite{Balents1996,Seidel2005,Dolfi2015,Dodaro2017,Jiang2018,Jiang2019HC,Jiang2019YF} For instance, recent DMRG calculations have found evidences for the LE liquid ground state of the lightly doped Hubbard model, where the undoped system is not a QSL, on a square lattice four-leg cylinder.\cite{Jiang2019HC,Jiang2019YF} It has one gapless charge mode but a finite gap in the spin sector. Although both the charge-density-wave (CDW) and superconducting (SC) pair-field correlations decay with a power-law in the LE liquid, a state with \textit{dominant} SC rather than CDW correlations provides much stronger evidences that long-range superconductivity would exist in the 2D limit. This is indeed the case that we find in this study as explained below.

The spin-1/2 antiferromagnet on the triangular lattice (depicted in Fig.\ref{Fig:Lattice}) with nearest-neighbor (NN) and next-nearest-neighbor (NNN) Heisenberg interactions, i.e., the second term in Eq.(\ref{Eq:Ham}), is highly frustrated. A number of numerical simulations have provided strong evidence that its ground state is a QSL in the region of $0.07\leq J_2/J_1\leq 0.15$,\cite{Zhu2015,Hu2015,Zheng2015,Saadatmand2016,Gong2017,Gong2019,Iqbal2016,Hu2019} which hence can serve as an ideal platform to investigate the consequence of doping a QSL. Although it is still under debate whether this QSL is gapped or gapless in the 2D limit, it is consistent with a gapped (possibly ``$Z_2$"\cite{Moessner2001}) spin liquid on finite width cylinders. This is evidenced by the observed non-zero spin-gap and exponentially falling spin-spin correlations where the correlation length is significantly shorter than the either dimension of the cylinders, and short-range dimer-dimer and chiral-chiral correlations.\cite{Zhu2015,Hu2015,Zheng2015,Saadatmand2016,Gong2017} This is also consistent with our recent results by keeping an unprecedentedly large number of states on wider cylinders, where we find a gapped spin liquid ground state.\cite{Jiang2019SL} Therefore, these leave little doubt that doping this state corresponds to a doped gapped QSL.

\textbf{Model and Method:} %
In this paper, we employ DMRG\cite{White1992} 
to study the ground state properties of the hole-doped $t$-$J$ model on the triangular lattice, defined by the Hamiltonian%
\begin{eqnarray}\label{Eq:Ham}
H=-\sum_{ij\sigma} t_{ij} \left(\hat{c}^+_{i\sigma} \hat{c}_{j\sigma} + h.c.\right) + \sum_{ij}J_{ij}\left (\vec{S}_i\cdot \vec{S}_j - \frac{\hat{n}_i \hat{n}_j}{4} \right ).
\end{eqnarray}
Here $\hat{c}^+_{i\sigma}$ ($\hat{c}_{i\sigma}$) is the electron creation (annihilation) operator on site $i=(x_i,y_i)$ with spin $\sigma$. $\vec{S}_i$ is the spin operator and $\hat{n}_i=\sum_{\sigma}\hat{c}^+_{i\sigma}\hat{c}_{i\sigma}$ is the electron number operator. The electron hopping amplitude $t_{ij}$ is equal to $t_1$ ($t_2$) if $i$ and $j$ are NN (NNN) sites as shown in Fig. \ref{Fig:Lattice}. $J_1$ and $J_2$ are the spin superexchange interactions between NN and NNN sites, respectively. The Hilbert space is constrained by the no-double occupancy condition, $n_i\leq 1$. At half-filling, i.e., $n_i=1$, Eq.(\ref{Eq:Ham}) reduces to the spin-1/2 antiferromagnetic $J_1$-$J_2$ Heisenberg model.

The lattice geometry used in our simulations is depicted in Fig.\ref{Fig:Lattice}, where $\mathbf{e}_x=(1,0)$ and $\mathbf{e}_y=(1/2,\sqrt{3}/2)$ denote the two basis vectors. We take the lattice geometry to be cylindrical with periodic (open) boundary condition in the $\mathbf{e}_y$ ($\mathbf{e}_x$) direction. Here, we focus on cylinders with width $L_y$ and length $L_x$, where $L_x$ and $L_y$ are number of sites along the $\mathbf{e}_x$ and $\mathbf{e}_y$ directions, respectively. The total number of sites is $N=L_x\times L_y$ with $N_e=N$ electrons at half-filling. The doping level of the system is defined as $\delta=N_h/N$, where $N_h=N-N_e$ is the number of doped holes.

For the present study, we focus on the lightly doped case with $\delta\leq 1/12$ on width $L_y=4$ cylinders of length up to $L_x=144$ and width $L_y=6$ cylinders of length up to $L_x=72$. We set $J_1$=1 as an energy unit and $J_2=0.11$ such that the system is deep in the QSL phase at half-filling.\cite{Zhu2015,Hu2015,Zheng2015,Saadatmand2016,Gong2017,Gong2019} We consider $t_1=3$ and $t_2=t_1\sqrt{J_2/J_1}=0.995$ to make a connection to the corresponding Hubbard model.\cite{Misumi2017} The results also hold for other $t_1$ and $t_2$. We perform up to 100 sweeps and keep up to $m=12000$ states for width $L_y=4$ cylinders with a typical truncation error $\epsilon< 3\times 10^{-7}$, and up to $m=25000$ states for width $L_y=6$ cylinders with a typical truncation error $\epsilon< 4\times 10^{-6}$. This leads to excellent convergence for our results when extrapolated to $m=\infty$ limit. Further details of the numerical simulation are provided in the Supplemental Material (SM).

\begin{figure}
  \includegraphics[width=\linewidth]{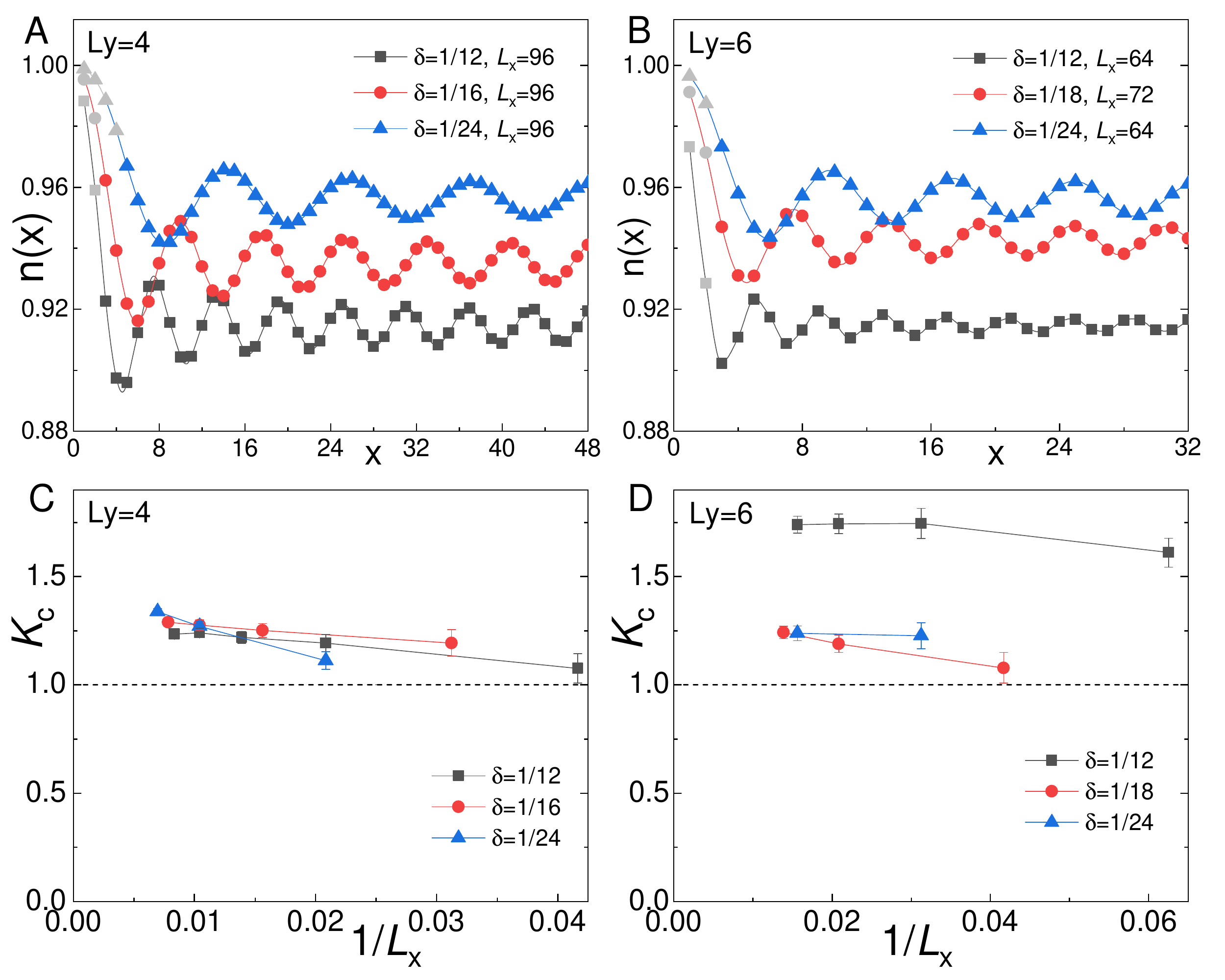}
  \caption{Charge density profiles and Luttinger exponent $K_c$. Charge density profiles $n(x)$ for (A) $L_y=4$ and (B) $L_y=6$ cylinders of length $L_x$ at various doping levels $\delta$. The extracted Luttinger exponent $K_c$ using Eq.(\ref{Eq:Kc}) for (C) $L_y=4$ and (D) $L_y=6$ cylinders as a function of $1/L_x$ and $\delta$, where the grey data points are excluded to minimize the boundary effect. Error bars denote the numerical uncertainty.}\label{Fig:NiKc}
\end{figure}

\textbf{Principal results:} %
We find that the system exhibits power-law CDW correlations corresponding to a local pattern of ``partial-filled" charge stripes. For width $L_y=4$ cylinders, the wavelength of the CDW, i.e., the spacings between two adjacent charge stripes, is $\lambda=1/2\delta$ corresponding to an ordering wavevector $Q=4\pi\delta$ with half a doped hole per unit cell (Fig.\ref{Fig:NiKc}A). This is similar with the lightly doped Hubbard model on a four-leg cylinder on the square lattice when the NNN electron hopping $t^\prime$ is included. For width $L_y=6$ cylinders, the CDW has a wavelength of $\lambda=1/3\delta$ and an ordering wavevector $Q=6\pi\delta$ with only one third of a doped hole per unit cell (Fig.\ref{Fig:NiKc}B).

\begin{figure}
  \includegraphics[width=\linewidth]{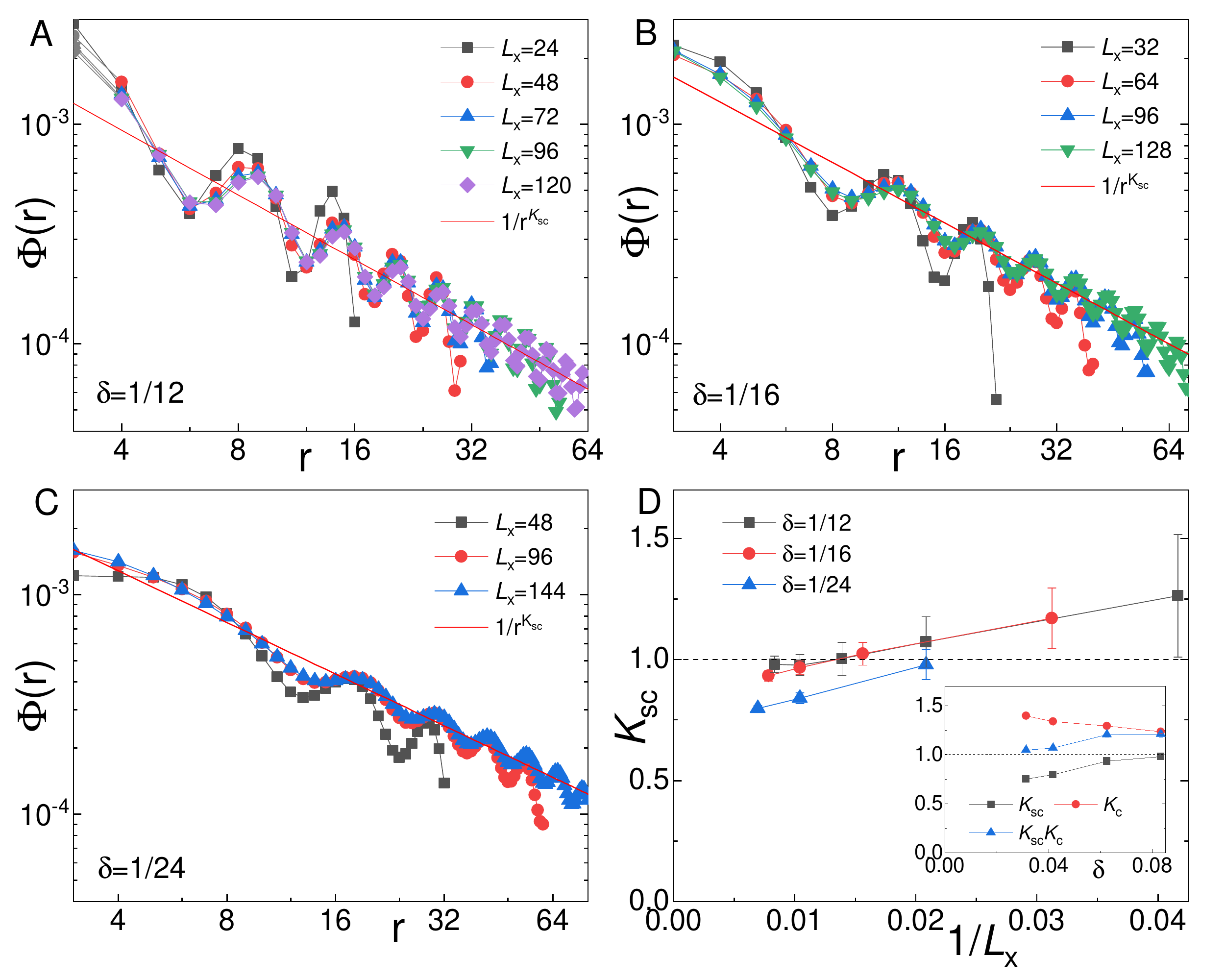}
  \caption{Superconducting correlations on $L_y=4$ cylinders. SC correlation functions $\Phi_{aa}(r)$ on $L_y=4$ cylinders of length $L_x$ for (A) $\delta=1/12$, (B) $\delta=1/16$ and (C) $\delta=1/24$ on double-logarithmic scales, where $r$ is the distance between two Cooper pairs in the $\mathbf{e}_x$ direction. The solid lines denote power-law fitting $\Phi_{aa}(r)\sim r^{-K_{sc}}$. (D) Luttinger exponent $K_{sc}$ as a function of $\delta$ and $1/L_x$. Inset: The extracted exponents $K_{sc}$, $K_c$ and their product $K_{sc}K_c$ from longest available cylinders as a function of $\delta$. Error bars denote the numerical uncertainty.}\label{Fig:Ly4SC}
\end{figure}

\begin{figure}
  \includegraphics[width=\linewidth]{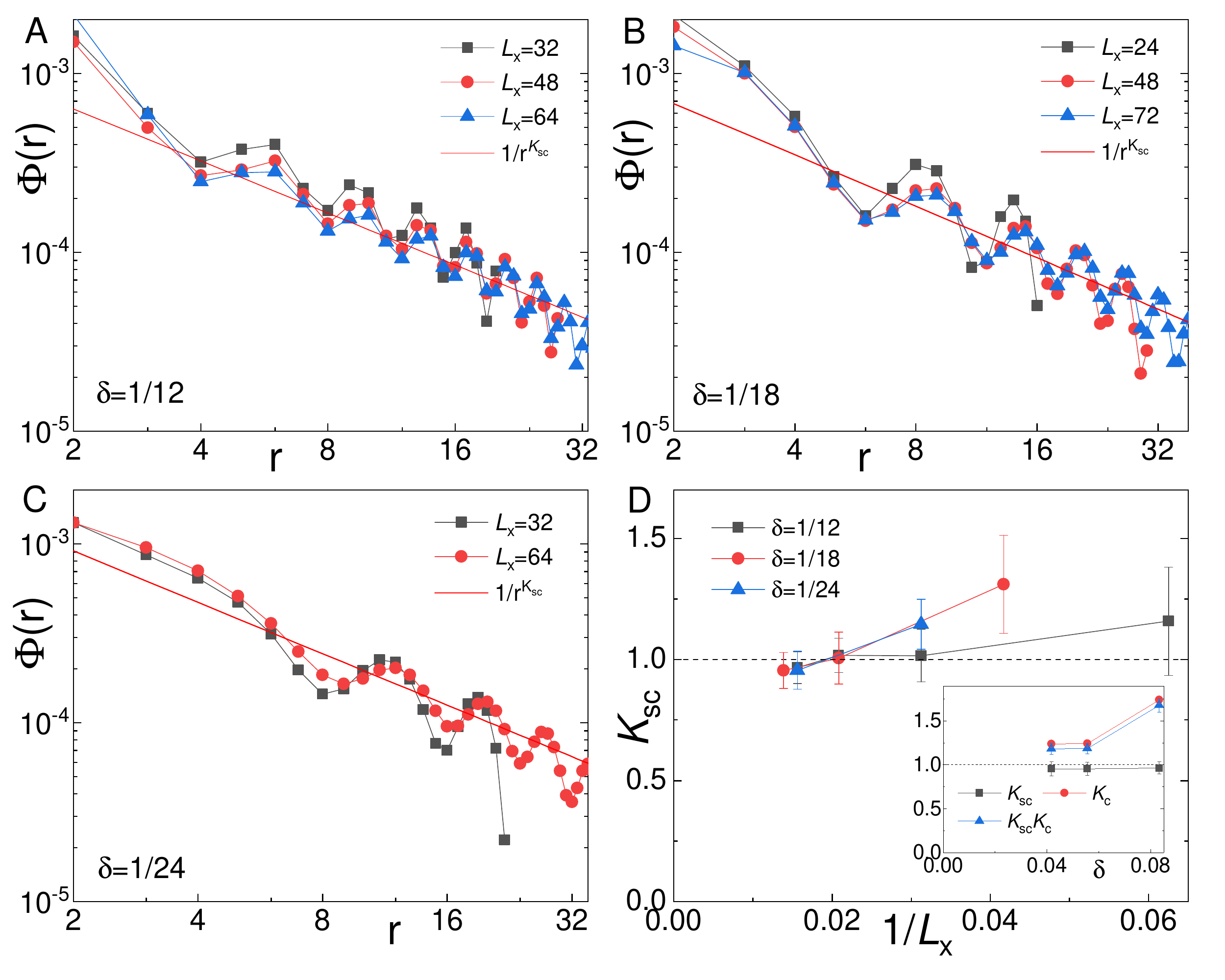}
  \caption{Superconducting correlations on $L_y=6$ cylinders. SC correlation functions $\Phi_{aa}(r)$ on $L_y=6$ cylinders of length $L_x$ for (A) $\delta=1/12$, (B) $\delta=1/18$ and (C) $\delta=1/24$ on double-logarithmic scales, where $r$ is the distance between two Cooper pairs in the $\mathbf{e}_x$ direction. The solid lines denote power-law fitting $\Phi_{aa}(r)\sim r^{-K_{sc}}$. (D) Luttinger exponent $K_{sc}$ as a function of $\delta$ and $1/L_x$. Inset: The extracted exponents $K_{sc}$, $K_c$ and their product $K_{sc}K_c$ from longest available cylinders as a function of $\delta$. Error bars denote the numerical uncertainty.} \label{Fig:Ly6SC}
\end{figure}

\begin{figure}
  \includegraphics[width=\linewidth]{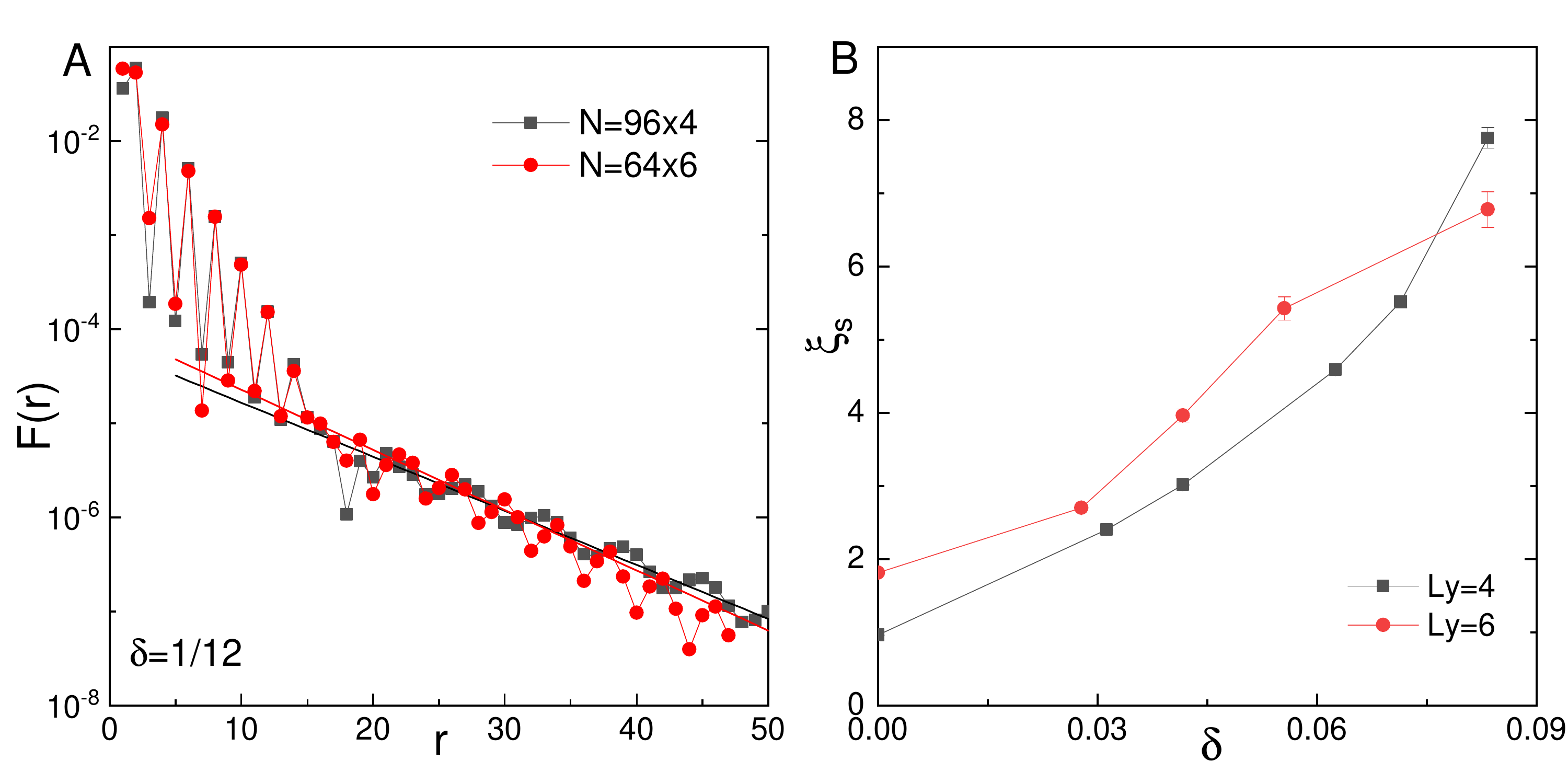}
  \caption{Spin-spin correlations. Examples of spin-spin correlation functions $F(r)$ at doping level $\delta=1/12$ for (A) $L_y=4$ and $L_y=6$ cylinders on the semi-logarithmic scale. Solid lines denote exponential fitting $F(r)\sim e^{-r/\xi_s}$, where $r$ is the distance between two sites in the $\mathbf{e}_x$ direction. (B) The spin-spin correlation length $\xi_s$ as a function of $\delta$ for both $L_y=4$ and $L_y=6$ cylinders. Error bars denote the numerical uncertainty.}\label{Fig:SS}
\end{figure}

\begin{figure}
  \includegraphics[width=\linewidth]{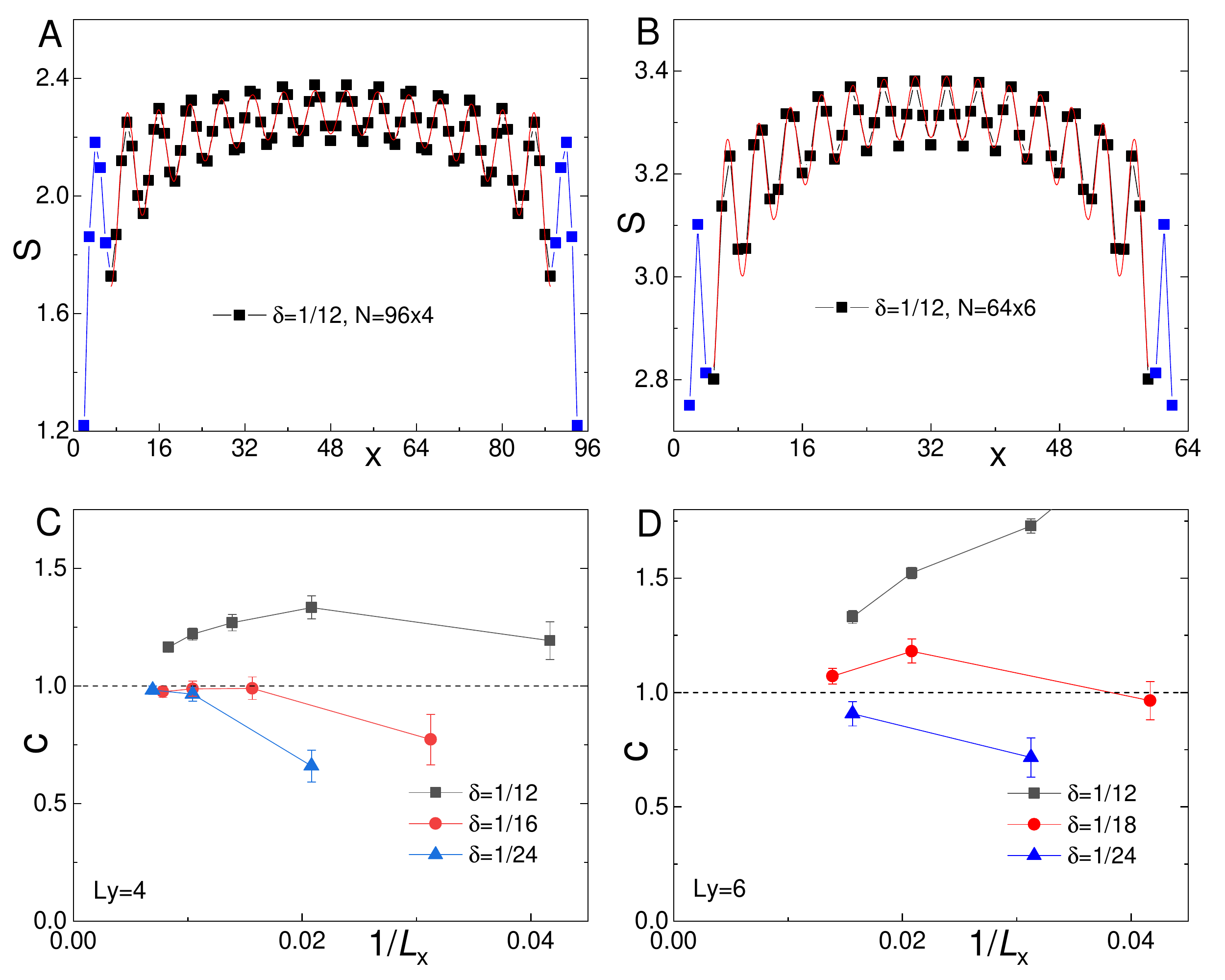}
  \caption{Von Neumann entanglement entropy $S$. $S$ is shown at doping level $\delta=1/12$ for (A) $L_y=4$ and (B) $L_y=6$ cylinders. Extracted central charge $c$ using Eq.(\ref{Eq:EE}) for (C) $L_y=4$ and (D) $L_y=6$ cylinders a a function of $L_x$ and $\delta$. Note that a few data points in blue from each end are removed to minimize boundary effects. Error bars denote the numerical uncertainty.}\label{Fig:Entropy}
\end{figure}

Because $L_x\gg L_y$, our system can be thought of as 1D, and we find that the ground state of the system is consistent with that of a LE liquid\cite{Luther1974}, which is characterized by gapless charge excitations (Fig.\ref{Fig:NiKc}, \ref{Fig:Ly4SC} and \ref{Fig:Ly6SC}) but a gap in the spin sector (Fig.\ref{Fig:SS}), with one gapless charge mode (Fig.\ref{Fig:Entropy}). At long distance, the spatial decay of the CDW correlations are dominated by a power-law with the Luttinger exponent $K_c$. The exponent $K_c$ can be obtained by fitting the charge density oscillations (Friedel oscillations) induced by the boundaries of the cylinder\cite{White2002}
\begin{eqnarray}
n(x)=n_0 + \delta n\ast {\rm cos}(2k_F x + \phi) x^{-K_c/2}.\label{Eq:Kc}
\end{eqnarray}
Here $n(x)=\langle \hat{n}(x)\rangle$, where $\hat{n}(x)=\sum_{y=1}^{L_y}\hat{n}(x,y)/L_y$ is the local rung density operator and $x$ is the rung index of the cylinder. $\delta n$ is a non-universal amplitude, $\phi$ is a phase shift, $n_0$ is the background density and $k_F$ is the Fermi wavevector. Note that a few data points (Fig.\ref{Fig:NiKc}A and B, grey color) are excluded to minimize the boundary effect and improve the fitting quality. The extracted exponent $K_c>1$ on both width $L_y=4$ and $L_y=6$ cylinders at all doping levels, in particular, $\delta=1/12$ on width $L_y=6$ cylinders (Fig.\ref{Fig:NiKc}C and D). Similarly, $K_c$ can also be obtained from the charge density-density correlation function which gives consistent results (see SM for details).

At long distance, the SC pair-field correlation $\Phi(r)$, defined in Eq.(\ref{Eq:SC}), is also characterized by a power-law with the appropriate Luttinger exponent $K_{sc}$ defined by%
\begin{eqnarray}
\Phi(r)\propto r^{-K_{sc}},\label{Eq:Ksc}
\end{eqnarray}
where $r$ is the distance between two Cooper pairs along the $\mathbf{e}_x$ direction. The extracted exponent is shown in Fig.\ref{Fig:Ly4SC}D and Fig.\ref{Fig:Ly6SC}D, where we can see clearly that $K_{sc}<1$ when $L_x\gg L_y$ for both $L_y=4$ and $L_y=6$ cylinders. As expected theoretically from the LE liquid\cite{Luther1974}, we find that the relation $K_c K_{sc}=1$ holds within the numerical uncertainty and finite-size effect. Although a notable deviation can be seen for the doping level $\delta=1/12$ on a width $L_y=6$ cylinders (Fig.\ref{Fig:Ly6SC}D inset), it can be attributed to the finite-size effect, which is also consistent with that of the central charge $c$ shown in Fig.\ref{Fig:Entropy}D. However, the apparent trend that $c$ approaches to the expected value $c=1$ in the limit $L_x\gg L_y$ shows that the ground state of the system is consistent with a LE liquid. In any case, the fact that $K_c>K_{sc}$ (Fig.\ref{Fig:Ly4SC}D and Fig.\ref{Fig:Ly6SC}D) undoubtedly demonstrates the dominance of the SC correlations $\Phi(r)$ on both $L_y=4$ and $L_y=6$ cylinders, which strongly suggests that long-range superconductivity would exist in the 2D limit.

\textbf{Charge density wave order:} %
To describe the charge density properties of the ground state of the system, we have calculated the charge density profile $n(x)$. Fig.\ref{Fig:NiKc}A shows the $n(x)$ for width $L_y=4$ cylinders, which is consistent with the ``half-filled" charge stripe of wavelength $\lambda=1/2\delta$, i.e., spacings between two adjacent stripes, for all doping concentration $\delta$ with half a doped hole per unit cell. The charge density profile $n(x)$ for width $L_y=6$ cylinders is shown in Fig.\ref{Fig:NiKc}B, which is however consistent with ``third-filled" charge stripes for all $\delta$ with only one third of a doped hole per unit cell. As a result, it has a shorter wavelength $\lambda=1/3\delta$ than the ``half-filled" stripes at the same doping concentration $\delta$.

At long distances, we find that the CDW correlations decay with a power-law with the Luttinger exponent $K_c$. The exponent $K_c$, which was obtained by fitting the data points using Eq.(\ref{Eq:Kc}), is shown in Fig.\ref{Fig:NiKc}C for width $L_y=4$ cylinders and Fig.\ref{Fig:NiKc}D for width $L_y=6$ cylinders as a function of $L_x$ and $\delta$. For all cases, we find that $K_c$ increases with $L_x$, which is consistent with weaker CDW correlations in the longer cylinders. A notable difference between the $L_y=4$ and $L_y=6$ cylinders is the doping dependence of $K_c$ as shown in Fig.\ref{Fig:Ly4SC}D and Fig.\ref{Fig:Ly6SC}D. On the $L_y=4$ cylinders, the value of $K_c$ decreases with the increase of $\delta$ on width $L_y=4$ cylinder. However, $K_c$ increases with the increase of $\delta$ on width $L_y=6$ cylinders. In any case, our results show that $K_c>1$, or more precisely $K_c>1.2$ when $L_x\gg L_y$, for all doping levels on both $L_y=4$ and $L_y=6$ cylinders. Further details can be found in the SM.

\textbf{Superconducting correlation:} %
To test the possibility of superconductivity, we have also calculated the equal-time SC pair-field correlations. As the ground state of the system with even number of doped holes is always found to have spin 0, we focus on spin-singlet pairing.\cite{SM} A diagnostic of the SC order is the pair-field correlator, defined as%
\begin{eqnarray}
\Phi_{\alpha\beta}(r)=\frac{1}{L_y}\sum_{y=1}^{L_y}\ \langle \Delta_\alpha^\dagger (x_0,y)\Delta_\beta(x_0+r,y)\rangle.\label{Eq:SC}
\end{eqnarray}
Here $\Delta_\alpha^\dagger(x,y)=\frac{1}{\sqrt{2}}[c_{(x,y),\uparrow}^\dagger c_{(x,y)+\alpha,\downarrow}^\dagger - c_{(x,y),\downarrow}^\dagger, c_{(x,y)+\alpha,\uparrow}^\dagger]$ is the spin-singlet pair-field creation operator, where the bond orientations are designated $\alpha=$a, b and c (Fig.\ref{Fig:Lattice}), $(x_0,y)$ is the reference bond with $x_0\sim L_x/4$, and $r$ is the distance between two bonds in the $\mathbf{e}_x$ direction.

Due to the presence of CDW modulations (Fig.\ref{Fig:NiKc}), SC correlations $\Phi_{\alpha\beta}(r)$ exhibit similar spatial oscillations with $n(x)$. For a given cylinder of length $L_x$, we determine the decay of SC correlations with reference bond located at the peak position around $x_0\sim L_x/4$ of the charge density distribution $n(x)$ to minimize the boundary effect\cite{Jiang2018,Jiang2019HC}. Examples of $\Phi_{aa}(r)$ are given in Fig.\ref{Fig:Ly4SC} and Fig.\ref{Fig:Ly6SC} for width $L_y=4$ and $L_y=6$ cylinders, respectively. For a given cylinder of length $L_x$, we extrapolate $\Phi_{aa}(r)$ to the limit $\epsilon=0$ using a second-order polynomial function with four data points of largest number of states, typically $m=6000\sim 12000$ for width $L_y=4$ cylinders and $m=15000\sim 25000$ for width $L_y=6$ cylinders. This gives accurate values of $\Phi_{aa}(r)$ for reliable finite-size scaling. Further details are provided in the SM.

As the ground state of the system is a spin-singlet state, the pair symmetry of the SC correlations is consistent with either $s$-wave or $d$-type wave, including the topological $d+id$-wave and nematic $d$-wave which breaks the $C_3$ lattice rotational symmetry. In the following we show that the pair symmetry is only consistent with the nematic $d$-wave.\cite{Feiguin2008,Raghu2010} 
First of all, our results show that the ground state wavefunction of the system is purely real and the imaginary part of all the SC correlations $\Phi_{\alpha\beta}(r)$ is zero. As a result, the pair symmetry is inconsistent with $d+id$-wave. Second, the SC correlations change sign between different bonds, i.e., $\Phi_{ab}\sim \Phi_{ac}\sim -0.4\Phi_{aa}$ (see SM for details), so the pair symmetry is also inconsistent with $s$-wave. Consequently, the remaining option is the nematic $d$-wave, which is indeed consistent with our results. The dominance of the SC correlations on $a$ bonds than other bonds, i.e., $\Phi_{bb}\sim \Phi_{cc}\sim 0.2\Phi_{aa}$ on both the $L_y=4$ and $L_y=6$ cylinders, demonstrate the robustness of the nematic order. Therefore, we conclude that the SC pair symmetry is nematic $d$-wave, where the $C_3$ lattice rotational symmetry is broken.


Similar to the CDW correlations, the SC correlations $\Phi(r)$ of the system on both width $L_y=4$ (Fig.\ref{Fig:Ly4SC}) and $L_y=6$ cylinders (Fig.\ref{Fig:Ly6SC}) also decay with a power law, whose exponent $K_{sc}$ was obtained by fitting the results using Eq.(\ref{Eq:Ksc}).\cite{Jiang2018,Jiang2019HC,Jiang2019YF} The fact that the observed $K_{sc}<1$ holds for all the cases when $L_x\gg L_y$ undoubtedly demonstrates the dominance of the SC correlations over the CDW correlations since $K_c>1>K_{sc}$. These findings are in stark contrast to the case of doping QSL on the Kagome lattice\cite{Jiang2017}, where the SC correlations decay exponentially but the CDW correlations have 2D long-range order. This indicates that the lattice geometry may play a critical role in giving rise to superconductivity in doping a QSL.

\textbf{Spin-spin correlation:} %
To describe the magnetic properties of the ground state, we calculate the spin-spin correlation functions defined as%
\begin{eqnarray}\label{Eq:SpinCor}
F(r)=\frac{1}{L_y}\sum_{y=1}^{L_y}|\langle \vec{S}_{x_0,y}\cdot \vec{S}_{x_0+r,y}\rangle |, 
\end{eqnarray}
where $\vec{S}_{x,y}$ is the spin operator on site $i=(x,y)$. $(x_0,y)$ is the reference site with $x_0\sim L_x/4$ and $r$ is the distance between two sites in the $\mathbf{e}_x$ direction. Following similar procedure as for $n(x)$ and $\Phi(r)$, we first extrapolate $F(r)$ to the limit $\epsilon\rightarrow 0$ for a given cylinder of length $L_x$ and then perform finite-size scaling as a function of $L_x$. As shown in Fig.\ref{Fig:SS}A, $F(r)$ decays exponentially as $F(r)\sim e^{-r/\xi_s}$, with a correlation length
$\xi_s=1\sim 8$ lattice spacings for the $L_y=4$ cylinders and $\xi_s=2\sim 7$ lattice spacings for the $L_y=6$ cylinders (Fig.\ref{Fig:SS}B) for doping concentration $0\leq \delta\leq 1/12$. Therefore, the spin-spin correlations are short-ranged with a finite spin gap, which is similar with the QSL at half-filling.

\textbf{Central charge:} %
A key feature of the LE liquid is that it has a single gapless charge mode with central charge $c=1$, which can be obtained by calculating the von Neumann entropy $S=-\rm Tr \rho ln \rho$, where $\rho$ is the reduced density matrix of a subsystem with length $x$. For critical systems in 1+1 dimensions described by a conformal field theory, it has been established\cite{Calabrese2004,Fagotti2011} that for an open system of length $L_x$,%
\begin{eqnarray}
S(x)&=&\frac{c}{6} \ln \big[\frac{4(L_x+1)}{\pi} \sin \frac{\pi(2x+1)}{2(L_x+1)}|\sin k_F|\big] \nonumber \\
&+&\tilde{A} \frac{\sin[k_F(2x+1)]}{\frac{4(L_x+1)}{\pi} \sin \frac{\pi(2x+1)}{2(L_x+1)}|\sin k_F|}+ \tilde{S},\label{Eq:EE}
\end{eqnarray}
where $\tilde{A}$ and $\tilde{S}$ are model dependent fitting parameters, and $k_F$ is the Fermi momentum. Examples of $S(x)$ are shown in Fig.\ref{Fig:Entropy}A and B for the $L_y=4$ and $L_y=6$ cylinders, respectively. For a given cylinder of length $L_x$, a few data points in blue close to the ends are excluded in the fitting to minimize the boundary effect. The extracted central charge $c$ is shown in Fig.\ref{Fig:Entropy}C and D as a function of $L_x$ and $\delta$. Although the value of $c$ deviates notably from $c=1$ on short cylinders, apparently it approaches $c=1$ with the increase of $L_x$ by taking into account the numerical uncertainty and finite-size effects. The result is consistent with one gapless charge mode with $c=1$, which provides additional evidence for the presence of the LE liquid in doping the QSL on the triangular lattice.

\textbf{Summary and discussion:} %
Taken together, our results show that the ground state of a doped QSL on the triangular lattice is consistent with a LE liquid with a finite gap in the spin sector. Although both CDW and SC correlations decay with a power-law, the fact that $K_c>1>K_{sc}$ on both the $4$-leg and $6$-leg cylinders when $L_x\gg L_y$ at all doping levels undoubtedly demonstrates the dominance of the SC correlations with strong divergent SC susceptibility. To the best of our knowledge, this is the first numerical realization of dominant superconductivity in a doped QSL, which provides strong evidence that doping a QSL would naturally give rise to long-range superconductivity in the 2D limit.\cite{Anderson1987,Kivelson1987,Rokhsar1988,Laughlin1988} In this paper, we have focused on the lightly doped case, it will also be interesting to study the higher doping case as well as the consequence of doping different phases on the triangular lattice such as the magnetic ordered phase. Answering these questions may lead to better understanding of the mechanism of high temperature superconductivity.

\textbf{Acknowledgement:}
We are grateful to S. Kivelson, Z. Wang and T. Devereaux for insightful discussions and suggestions to improve the manuscript and S. Kivelson for pointing out the nematic $d$-wave symmetry. We are also grateful to D. J. Scalapino, Z. Y. Weng, H. Yao, Y. F. Jiang, S. Raghu for insightful discussions. This work was supported by the Department of Energy, Office of Science, Basic Energy Sciences, Materials Sciences and Engineering Division, under Contract DE-AC02-76SF00515. Parts of the computing for this project was performed on the Sherlock cluster.


\appendix 

\begin{center}
\noindent {\large {\bf Supplemental Material}}
\end{center}

\renewcommand{\thefigure}{S\arabic{figure}}
\setcounter{figure}{0}
\renewcommand{\theequation}{S\arabic{equation}}
\setcounter{equation}{0}

\section{I. Numerical convergence}\label{SM:Convergence} %
We have checked the numerical convergence of our DMRG simulations regarding various symmetries such as spin rotational symmetry. It is known that the ground state of a finite system in one or two dimensions cannot spontaneously break any continuous symmetry. Therefore, the true ground state of the $t$-$J$ model on finite cylinders should preserve the $SU(2)$ spin rotational symmetry. This can be considered as one of the key signatures to determine whether a DMRG simulation has converged to the true ground state. We take two routes to address this issue in our DMRG simulation. First of all, we calculate the expectation value of the $z$-component of the spin operator, i.e., $\langle\hat{S}^z_i\rangle$, where $i$ labels the lattice site. As the true ground state is an equal-weight superposition of $|S^z=1/2\rangle$ and $|S^z=-1/2\rangle$ spin states, then $\langle \hat{S}^z_i\rangle=0$ for all sites $i$. A simple measurement of this condition is to define a quantity $m_z=\sum_{i=1}^N |\langle S^z_i\rangle|/N$, which should vanish as the DMRG simulation converges to the true ground state. In all of our DMRG simulations for both $L_y=4$ and $L_y=6$ cylinders, we find that $m_z=0$ even when we keep a relatively small number of states, suggesting that our simulations have converged. Second, the $SU(2)$ spin rotational symmetry requires that the relation $\langle S^x_i S^x_j\rangle$=$\langle S^y_i S^y_j\rangle$=$\langle S^z_i S^z_j\rangle$ holds between two arbitrary sites $i$ and $j$, which is again fulfilled in our simulations. In addition to spin rotational symmetry, other symmetries including both the lattice translational symmetry in $\mathbf{e}_y$ direction and reflection symmetry in the $\mathbf{e}_x$ direction are also satisfied. Therefore, we conclude that our numerical simulation has converged to the true ground state.

\section{II. Further calculation details} %
To reliably describe ground state properties, we have explored the role of cylinder size and boundary effects. In the current study, we typically start with a random state in our calculation. However, to elucidate the reliability of our results, we also check our calculations by adding a pinning field with the appropriate wavelength to stabilize a CDW state. We find that in all the cases it is sufficient to add the pinning field during the initial sweeps of the calculation and ramp its amplitude to zero in a few subsequent sweeps. This happens only for the smallest number of states that we have considered, i.e., $m=2187$, while for the subsequent calculations with $m>2187$ it is not necessary to hold a finite (even vanishingly small) pinning field to stabilize the charge stripe pattern. This gives us the same results as we start from a completely random initial state without any pinning field, which undoubtedly demonstrates the reliability of our study. Moreover, it is not necessary to apply pinning pair-field to stabilize the superconductivity in our DMRG calculation.

\begin{figure}
  \includegraphics[width=\linewidth]{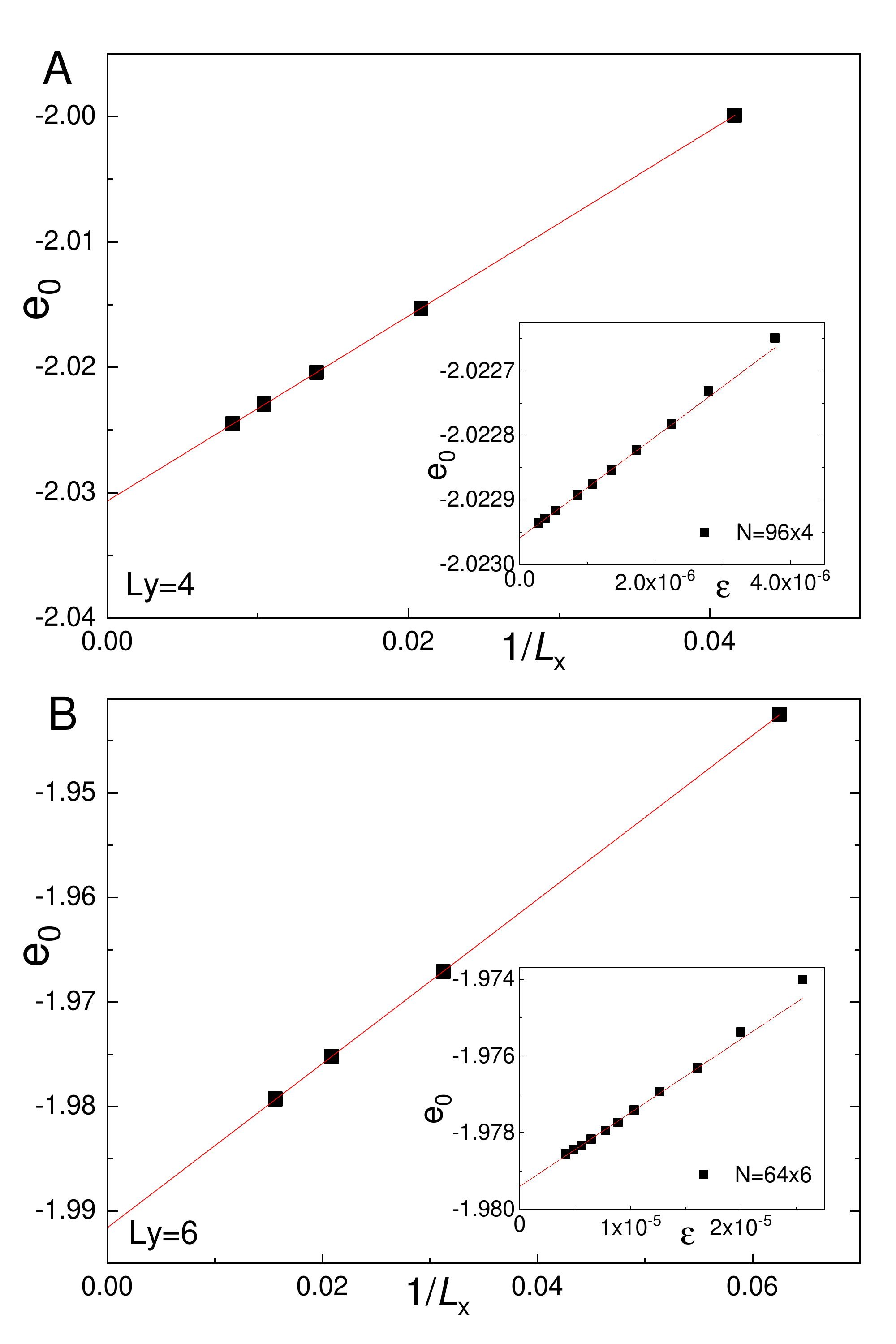}
  \caption{Ground state energy of the $t$-$J$ model. Ground state energy per site $e_0$ for $\delta=1/12$ as a function of $1/L_x$ for width (A) $L_y=4$ and (B) $L_y=6$ cylinders. Insets: Examples of truncation error $\epsilon$ extrapolation of $e_0$ for the (A) $N=96\times 4$ and (B) $N=64\times 6$ cylinders.The red lines show the extrapolation using a linear function.}\label{FigS:EG}
\end{figure}

\section{III. Ground state energy} %
Fig.\ref{FigS:EG} shows examples of truncation error $\epsilon$ extrapolation of the ground state energy per site $e_0=E_0/N$, where $E_0$ is the total energy of a system for $N=96\times 4$ and $N=64\times 6$ at doping level $\delta=1/12$. By keeping $m=2187\sim 12000$ number of states for the $L_y=4$ cylinders, and $m=2187\sim 25000$ number of states for the $L_y=6$ cylinders, we are able to converge to the true ground state of the system which preserves all the symmetries of the Hamiltonian, including the $SU(2)$ spin rotational symmetry, lattice translational symmetry in the $\mathbf{e}_y$ direction and reflection symmetry in the $\mathbf{e}_x$ direction. The truncation error extrapolation using a linear function with $m=3200\sim 12000$ for $N=96\times 4$ cylinder gives $e_0=-2.02296(1)$. For $N=64\times 6$ cylinder, the truncation error extrapolation using a linear function with $m=10000\sim 25000$ gives $e_0=-1.97929(2)$. The ground state energy $e_0$ for other cylinders and doping concentrations $\delta$ can be obtained similarly. Consequently, we can obtain accurate estimates of the ground state energies in the long cylinder limit, i.e., $L_x\rightarrow \infty$, by carrying out a finite-size scaling as a function of $1/L_x$. Examples of the extrapolation are shown in Fig.\ref{FigS:EG} for the (A) $L_y=4$ and (B) $L_y=6$ cylinders at $\delta=1/12$, in which all energies of various cylinder length $L_x$ fall perfectly onto a linear fit, with a linear regression $R^2=1$. This gives an energy $e_0=-2.03064(2)$ for the $L_y=4$ cylinder, and $e_0=-1.99158(4)$ for the $L_y=6$ cylinders in the limit $L_x=\infty$. Similarly, we have also obtained the ground state energies for other doping concentrations, e.g., $e_0=-1.86683(1)$ at $\delta=1/16$ and $e_0=-1.69974(1)$ at $\delta=1/24$ for the $L_y=4$ cylinders, and $e_0=-1.78442(1)$ at $\delta=1/18$ and $e_0=-1.67589(1)$ at $\delta=1/24$ for the $L_y=6$ cylinders.

\begin{figure}
  \includegraphics[width=\linewidth]{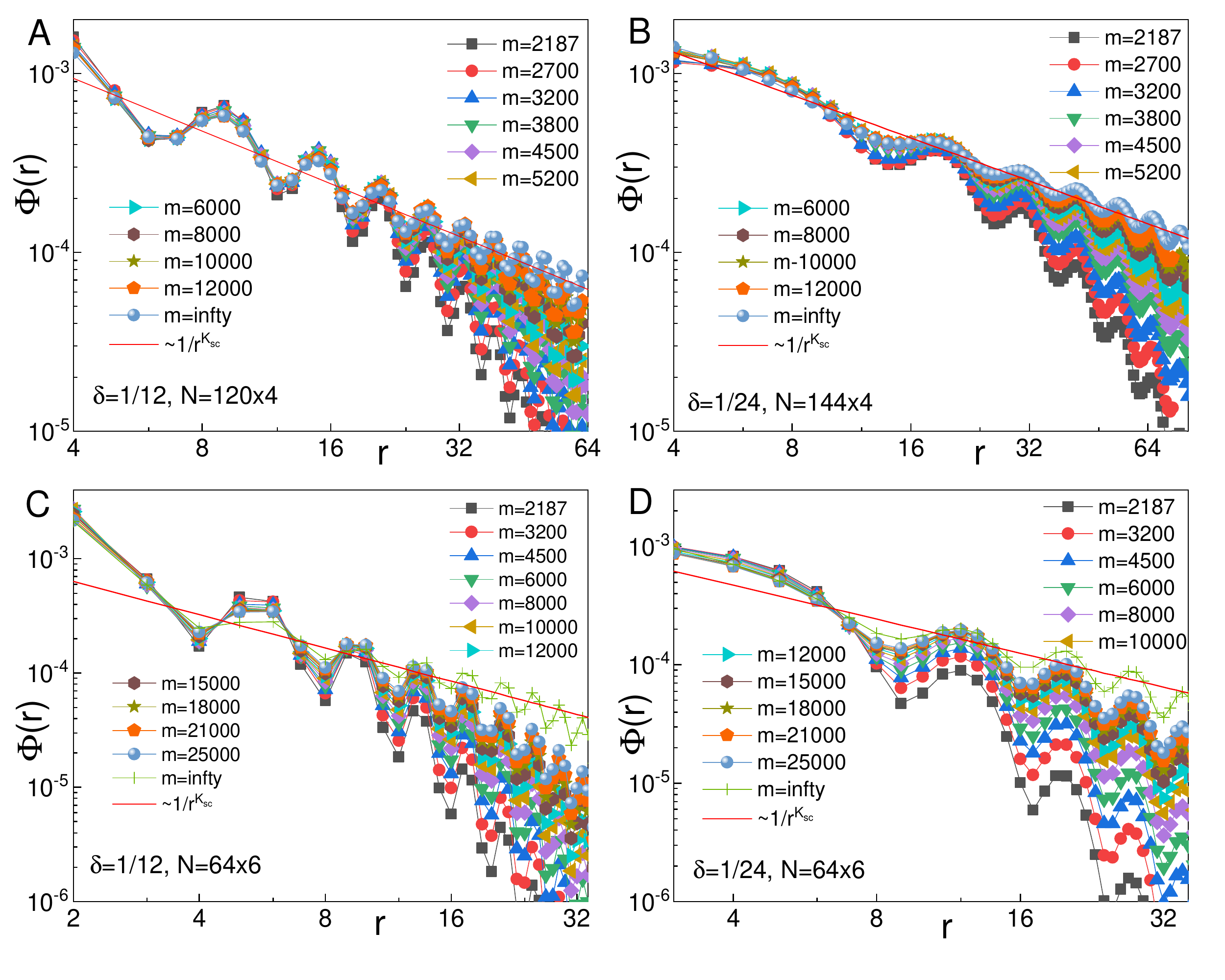}
  \caption{Convergence of the SC correlations. SC correlation $\Phi_{aa}(r)$ on the (A) $N=120\times 4$ cylinder at $\delta=1/12$ and (B) $N=144\times 4$ cylinder at $\delta=1/24$, by keeping $m=2187\sim 12000$ states, and the $N=64\times 6$ cylinders at (C) $\delta=1/12$ and (D) $\delta=1/24$, by keeping $m=2187\sim 25000$ states. All the figures are plotted in the double-logarithmic scales, where $r$ is the distance between two Cooper pairs in the $\mathbf{e}_x$ direction. The red lines label the power-law fit $\Phi_{aa}(r)\sim 1/r^{K_{sc}}$.}\label{FigS:SCCor}
\end{figure}

\begin{figure}
  \includegraphics[width=\linewidth]{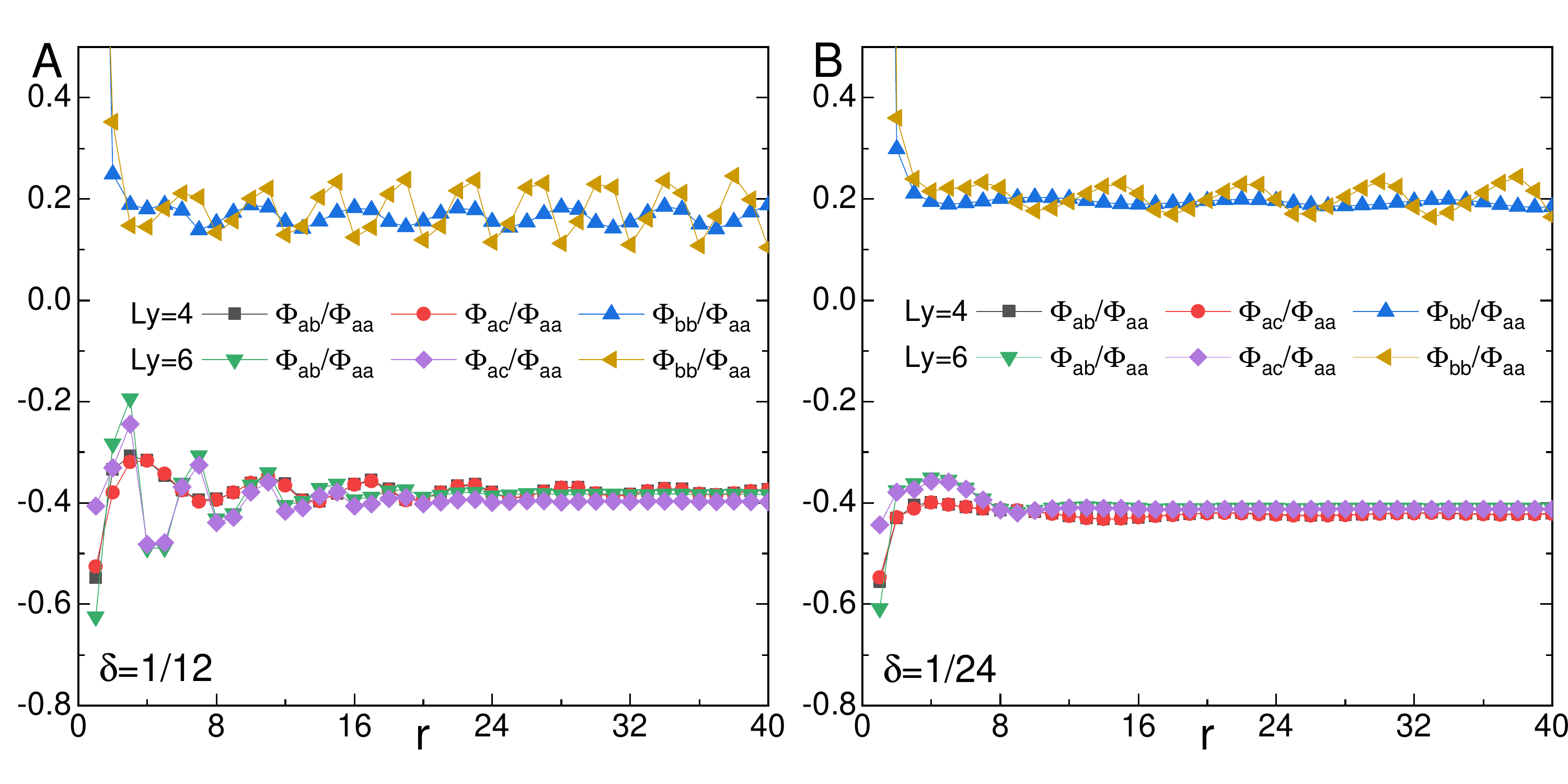}
  \caption{Pair symmetry of the SC correlations. Ratio of the SC correlations $\Phi_{ab}/\Phi_{aa}$, $\Phi_{ac}/\Phi_{aa}$ and $\Phi_{bc}/\Phi_{aa}$ on both the $L_y=4$ and $L_y=6$ cylinders at (A) $\delta=1/12$ and (B) $\delta=1/24$. $r$ is the distance between two Cooper pairs in the $\mathbf{e}_x$ direction.}\label{FigS:PairSym}
\end{figure}

\begin{figure}
  \includegraphics[width=\linewidth]{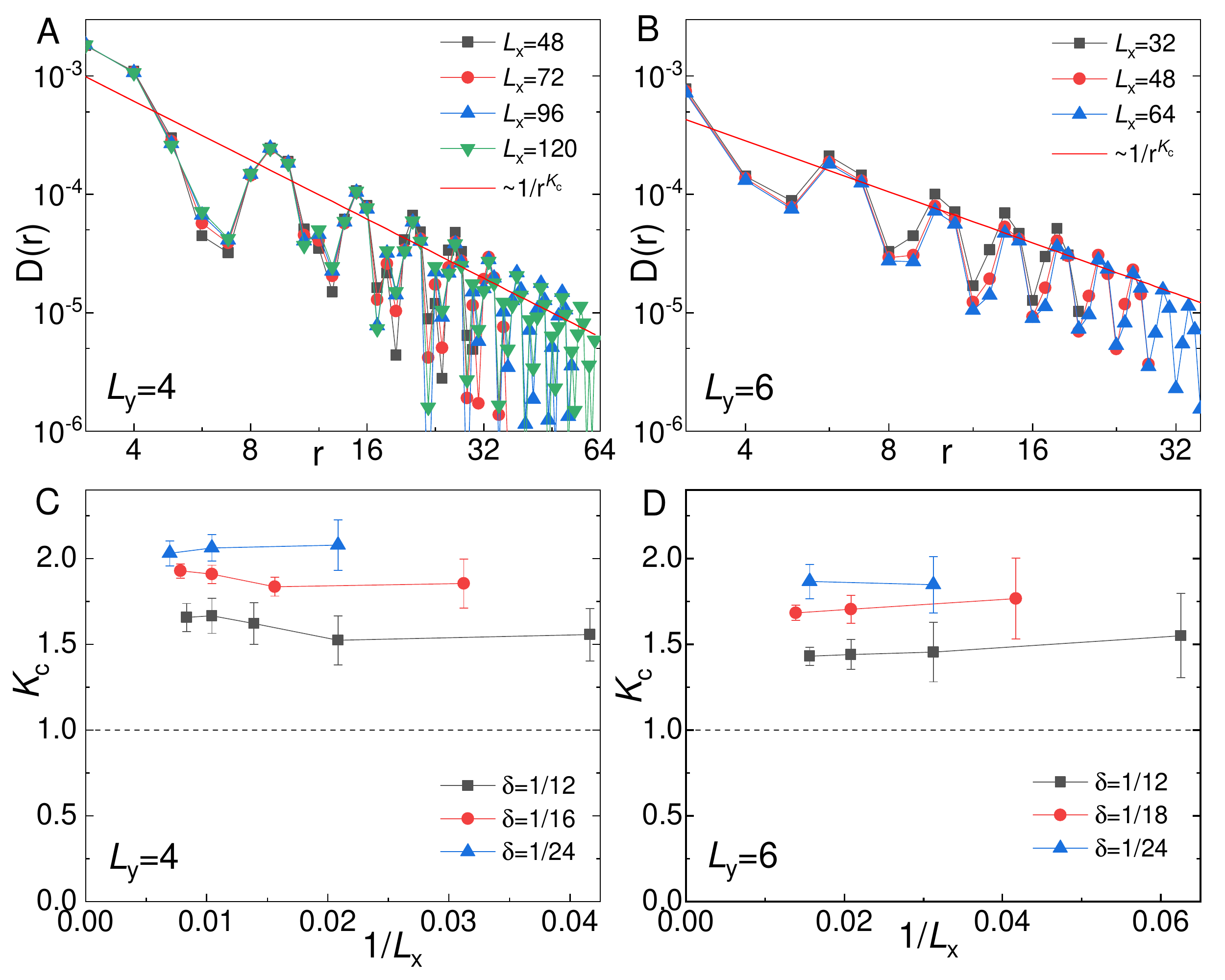}
  \caption{Charge density-density correlations. Charge density-density correlations $D(r)$ at $\delta=1/12$ for the (A) $L_y=4$ and (B) $L_y=6$ cylinders on double-logarithmic scales, where $r$ is the distance between two sites in the $\mathbf{e}_x$ direction. The solid lines denote a power-law fit $D(r)\sim r^{-K_c}$. Extracted exponent $K_c$ for the (C) $L_y=4$ and (D) $L_y=6$ cylinders as a function of $\delta$ and $1/L_x$. The dashed line is a guide for eyes and error bars denote the numerical uncertainty.}\label{FigS:DenCor}
\end{figure}

\section{IV. Convergence of superconducting correlations }\label{SM:SCCor} %
Fig.\ref{FigS:SCCor}A and B show the superconducting (SC) pair-field correlations $\Phi_{aa}(r)$ with $r=1\sim L_x/2$ in the $\mathbf{e}_x$ direction for width $L_y=4$ cylinders by keeping $m=2187\sim 12000$ states at doping levels $\delta=1/12$ and $\delta=1/24$. The blue circles label the extrapolated values to the limit $\epsilon=0$, i.e., $m=\infty$, using second-order polynomial function with four data points of largest number of states $m=6000\sim 12000$. To minimize the boundary and finite-size effect, the first few data points with small $r$ are excluded. As indicated by the red sold lines, the SC correlations are consistent with a power-law decay $\Phi_{aa}\propto r^{-K_{sc}}$ with $K_{sc}\sim 1$.

Fig.\ref{FigS:SCCor}C and D show the SC pair-field correlations $\Phi(r)_{aa}$ with $r=1\sim L_x/2$ in the $\mathbf{e}_x$ direction for width $L_y=6$ cylinders by keeping $m=2187\sim 25000$ states at doping levels $\delta=1/12$ and $\delta=1/24$. The green crosses label the extrapolated values to the limit $\epsilon=0$, i.e., $m=\infty$, using second-order polynomial function with four data points of the largest number of states $m=15000\sim 25000$. To minimize the boundary and finite-size effect, the first few data points with small $r$ are excluded. Similarly with the $L_y=4$ cylinders, $\Phi_{aa}(r)$ are also consistent with a power-law decay $\Phi_{aa}(r)\propto r^{-K_{sc}}$ with $K_{sc}\sim 1$, as indicated by the red solid lines.

In addition to the spin-singlet SC correlation, we have also calculated the spin-triplet SC correlation. However, it is much weaker than the spin-singlet SC correlation, suggesting that spin-triplet superconductivity is unlikely.

\section{V. Superconducting pair symmetry}\label{SM:PairSym} %
We have shown in the main text that the pair symmetry of the SC correlations is consistent with the nematic $d$-wave. Here, we provide more results to further support this. Fig.\ref{FigS:PairSym} shows examples of the SC correlations for both width $L_y=4$ and $L_y=6$ cylinders at $\delta=1/12$ and $\delta=1/24$. For all the cases, we find that the SC correlations have the following relations: $\Phi_{bb}\sim \Phi_{cc}\sim 0.2\Phi_{aa}$, $\Phi_{ab}\sim \Phi_{ac}\sim - 0.4 \Phi_{aa}$, and $\Phi_{bc}\sim 0.2\Phi_{aa}$. Similar relation also holds for other doping levels, such as $\delta=1/16$ and $\delta=1/18$ etc.

\section{VI. Charge density-density correlations}\label{SM:DenCor} %
In addition to the charge density oscillation (Friedel oscillation), the exponent $K_c$ can also be extracted from the charge density-density correlation, which is defined as $D(r)=\langle (\hat{n}(x_0)-\langle \hat{n}(x_0)\rangle)(\hat{n}(x_0+r)-\langle \hat{n}(x_0+r)\rangle)\rangle$. Here $x_0$ is the rung index of the reference site and $r$ is the distance between two sites in the $\mathbf{e}_x$ direction. Following similar procedure as for $n(x)$ and $\Phi(r)$, we extrapolate $D(r)$ to the limit $\epsilon\rightarrow 0$ for a given cylinder of length $L_x$ and then perform finite-size scaling as a function of $L_x$. As shown in Fig.\ref{FigS:DenCor}A and B for $\delta=1/12$, $D(r)$ decays with a power-law whose exponent $K_c$ was obtained by fitting the results using $D(r)\propto r^{-K_c}$. Similarly, the exponent $K_c$ can be obtained for other $\delta$, which are shown in Fig.\ref{FigS:DenCor}C and D for the $L_y=4$ and $L_y=6$ cylinders, respectively. Although the exponent fitted from $D(r)$ slightly deviates quantitatively from that extracted from the charge density oscillation $n(x)$, which is due to the fact that the calculation of the multiple-operator correlation is less accurate than the single operator measurement. However, they are qualitatively consistent with each other with $K_c>1$.

\end{document}